\documentclass[pra,aps,twocolumn,showpacs,superscriptaddress]{revtex4}
\usepackage{amsmath,graphicx,epsfig}

\begin{document}

\title{Gyroscopes based on nitrogen-vacancy centers in diamond} 

\author{M.\ P.\ Ledbetter}\email{micah.ledbetter@gmail.com}
\address{Department of Physics, University of California at
Berkeley, Berkeley, California 94720-7300, USA}

\author{K. Jensen}
\address{Department of Physics, University of California at
Berkeley, Berkeley, California 94720-7300, USA}

\author{R. Fischer}
\address{Department of Physics, Technion – Israel Institute of Technology, Haifa 32000, Israel}

\author{A. Jarmola}
\address{Department of Physics, University of California at
Berkeley, Berkeley, California 94720-7300, USA}

\author{D. Budker}\email{dbudker@gmail.com}
\address{Department of Physics, University of California at
Berkeley, Berkeley, California 94720-7300, USA}
\address{Nuclear Science Division, Lawrence Berkeley National
Laboratory, Berkeley CA 94720, USA}

\date{\today}


\begin{abstract}
We propose solid-state gyroscopes based on ensembles of negatively charged nitrogen-vacancy (${\rm NV^-}$) centers in diamond. In one scheme, rotation of the nitrogen-vacancy symmetry axis will induce Berry phase shifts in the ${\rm NV^{-}}$ electronic ground-state coherences proportional to the solid angle subtended by the symmetry axis. We estimate sensitivity in the range of $5\times10^{-3}~{\rm rad/s/\sqrt{Hz}}$ in a  1~${\rm mm^3}$ sensor volume using a simple Ramsey sequence.  Incorporating dynamical decoupling to suppress dipolar relaxation may yield sensitivity at the level of $10^{-5}~{\rm rad/s/\sqrt{Hz}}$.  With a modified Ramsey scheme, Berry phase shifts in the ${\rm ^{14}N}$ hyperfine sublevels would be employed.  The projected sensitivity is in the range of $10^{-5}~{\rm rad/s/\sqrt{Hz}}$, however the smaller gyromagnetic ratio reduces sensitivity to magnetic-field noise by several orders of magnitude.  Reaching $10^{-5}~{\rm rad/s/\sqrt{Hz}}$ would represent an order of magnitude improvement over other compact, solid-state gyroscope technologies.
\end{abstract}



\maketitle

Inertial sensors form the basis for navigation systems in both military and civilian applications.  Such devices also find use in gravity research, for example, in tests of  Einstein's equivalence principle \cite{Muller2010} or to search for the Lens-Thirring effect \cite{Jentsch2004}, a prediction from general relativity of ``frame-dragging" in the presence of a massive rotating body. Present state-of-the-art sensors used for navigation in the airline industry are based on the Sagnac effect in fiber-optic bundles, with sensitivity at the level of $2\times 10^{-8}~{\rm rad /s/\sqrt{Hz}}$ \cite{Sanders2002}.  The same effect in large-area (${\rm \approx m^2}$) ring lasers has yielded sensitivity at the level of $2\times 10^{-10}~{\rm rad/s/\sqrt{Hz}}$ \cite{Stedman2003}.  Gyroscopes based on cold atom interferometry have also demonstrated sensitivity in the range of $10^{-9}$ to $10^{-10}~{\rm rad /s/\sqrt{Hz}}$ \cite{Gustavson2000,Jentsch2004}.  Noble gas nuclear spins, which feature long coherence times, can also be used to form rotation sensors \cite{LittonPatent,Larsen}, with demonstrated sensitivity of about $3\times 10^{-7}~{\rm rad/s/\sqrt{Hz}}$.  Variations on this scheme yield projected sensitivity in the range of $10^{-10}$ ${\rm rad/s/\sqrt{Hz}}$ \cite{Kornack2005} in an active volume of several ${\rm cm^3}$.  This concept has been extended to miniaturized versions \cite{Lust2008}.  Finally, commercially available vibrating microelectromechanical systems (MEMS) gyroscopes achieve sensitivities on the order of ${\rm 2\times 10^{-4} ~rad/s/\sqrt{Hz}}$ \cite{Invensense}.
Except for the latter two cases, these sensors rely on large volumes or enclosed areas.


Here we propose applying geometric-phase effects in diamond to form gyroscopic sensors from high density ensembles of ${\rm NV^-}$ centers. The key advantages of ${\rm NV^{-}}$ based sensors are their potentially small dimensions and thermal robustness, with the ability to operate from cryogenic temperatures to 600-700 K \cite{Toyli2012}. Recent theoretical work discusses the possibility of observing large geometric phase shifts in single ${\rm NV^-}$ centers in diamond \cite{Maclaurin2012}. Similar ideas to those proposed here have been discussed in Ref. \cite{Cappellaro2012}. Geometric phase shifts, also known as Berry phase shifts \cite{Berry1984}, arise when the symmetry axis of a system is rotated about an axis that is not parallel to this axis.  This results in relative phase shifts of the magnetic sublevels. One encounters Berry phase shifts, for example, when a spin system evolves in time in the presence of an external magnetic field, which itself rotates in space in such a way that the field vector traces a nonzero solid angle \cite{Commins1991}. In the case of adiabatic evolution, an eigenstate with magnetic quantum number $m$ acquires a Berry phase (in addition to the phase due to evolution under the static Hamiltonian) that is given by the product of $m$ and the solid angle traced by the field, $d\Omega$.  If the quantization axis is determined by an electric field, a similar effect occurs on rotating the electric field \cite{Budkerbook}. This is the case for nuclear quadrupoles interacting with an electric-field gradient, and such an effect was used to demonstrate a ${\rm ^{131}Xe}$ nuclear-spin gyroscope \cite{Appelt1994}.


The ${\rm NV^{-}}$ center features a ground state with electronic spin $S=1$ which can be initialized, manipulated, and detected via convenient optical and microwave transitions.  Furthermore, the lifetime of the ground-state coherences can be quite long, on the order of 2 ms \cite{Balasubramanian2009} in ${\rm ^{13}C}$ depleted samples, even without applying dynamic decoupling sequences.   The analysis presented below indicates that sensitivities at the level of $\approx 10^{-5}~{\rm rad/s/\sqrt{Hz}}$ should be achievable in mm scale packages.  This falls short of the sensitivities demonstrated by  state-of-the-art techniques based on larger sensors, though it surpasses that of commercially available vibrating MEMS sensors.  Furthermore, operation as a three-axis sensor may be possible by monitoring coherences in  ${\rm NV^-}$ centers with different orientations with respect to the host lattice.

\begin{figure}
  \includegraphics[width=3.6 in]{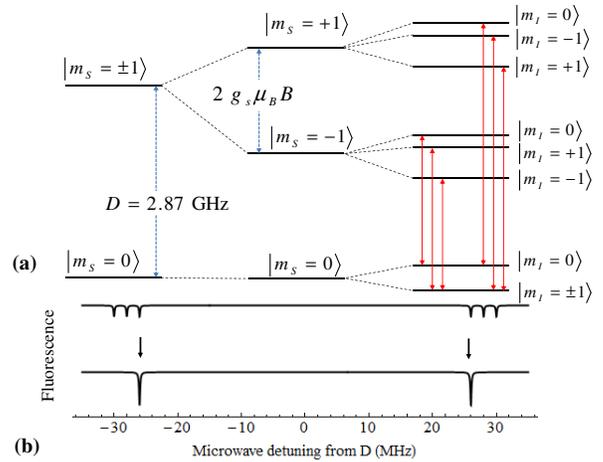}\\
  \caption{(a) Ground-state energy levels of NV$^-$ centers in the presence of a magnetic field of several G, and accounting for ${\rm ^{14}N}$ hyperfine coupling. The solid, double-headed red arrows show the allowed microwave transitions, which preserve the nitrogen spin, $\Delta m_I = 0$. (b) ${\rm ^{14}N}$ hyperfine resolved ODMR spectrum as a function of microwave detuning from $D$ for the case of unpolarized (upper trace) and polarized (lower trace) ${\rm ^{14}N}$ spins.}\label{Fig:14NHyperfine}
\end{figure}

The ground-state level structure of the NV$^-$ centers is shown in Fig.~\ref{Fig:14NHyperfine}(a).  Detailed discussions of the physics and spectroscopy of this defect can be found in Refs. \cite{Manson2006,Rogers2008,Doherty2012} and references therein. ${\rm NV^-}$ centers possess $C_{3v}$ symmetry, and can be oriented along any of the four crystallographic orientations.  The electronic ground state is spin-triplet $(S=1)$. In the absence of strain or external magnetic field, the $m_S = \pm 1$ levels are degenerate, where $m_S$ is the quantum number associated with the projection of $S$ along the symmetry axis. Due to the spin-spin interaction, the zero-field splitting between the $m_S = 0$ and $m_S = \pm 1$ levels is $D = 2.87~{\rm GHz}$.  Application of a magnetic field $B_z$ along the ${\rm NV^-}$ symmetry axis produces a shift of the ground-state energy levels according to $m_S g_s \mu_B B_z$, where $g_s\approx 2$ is the Land\'e factor for the electron and $\mu_B = 8.79~{\rm rad/s/G}$ is the Bohr magneton. (We focus here on a single orientation, assumed parallel to the magnetic field.  The projection of such a magnetic field onto the other three axes is suppressed by a factor of 3.  Therefore, for sufficiently large magnetic fields, $\approx 10~{\rm G}$, the microwave transitions of differently oriented ${\rm NV^-}$ centers are well isolated from ${\rm NV^-}$ centers whose symmetry axis is parallel to the field.  Since microwave transitions are excited resonantly, we can ignore the contributions from these other orientations.) In the presence of local strain, the $m_S = \pm 1$ sublevels are mixed and the corresponding eigenstates are split in energy by $2E$, where $E$ is the transverse zero-field splitting parameter, typically on the order of a few MHz for high ${\rm NV^-}$ density samples.  Application of magnetic fields of several gauss will overwhelm this, so that eigenstates are approximately those of $S_z$.  Finally, hyperfine coupling of the ${\rm ^{14}N}$ nuclear spin ($I = 1$) to the electronic degrees of freedom produces additional splittings as shown at the right of Fig. \ref{Fig:14NHyperfine}(a).

Optically detected magnetic resonance (ODMR) spectroscopy of the ground state can be accomplished as follows: Optical transitions from the $^3A_2$ ground states to the $^3E$ excited-state manifold are spin-conserving, preserving the magnetic quantum number $m_S$, and are typically excited via the phonon sideband (PSB) of the $^3A_2\rightarrow ^3E$ transition using 532 nm light.  Optical pumping of the $m_S=0$ ground state occurs through spin non-conserving decay pathways (due to spin-orbit coupling) involving singlet levels not shown in Fig. \ref{Fig:14NHyperfine}.  Fluorescence in the zero-phonon line (ZPL) at 637 nm and the associated PSB is brighter for ${\rm NV^-}$ starting in the $m_S=0$ state.  Thus, application of microwaves tuned to the $m_S = 0\rightarrow m_S = \pm 1$ ground-state transitions depopulates the $m_S = 0$ state and reduces fluorescence in the ZPL and its PSB.  Assuming the ${\rm NV^-}$ center is aligned with the magnetic field, and neglecting ${\rm ^{14}N}$ hyperfine coupling, fluorescence as a function of microwave frequency appears as a doublet, centered about $D$, with a splitting determined by the magnetic field $\Delta = 2 g_s\mu_B B_z$ (assuming $g_s\mu_B B_z \gg E$). If ${\rm ^{14}N}$ hyperfine coupling is resolved, each line of the doublet splits into a triplet.  ODMR spectra are schematically shown in Fig.~\ref{Fig:14NHyperfine}(b) for unpolarized and polarized ${\rm ^{14}N}$ spins (top and bottom traces, respectively).  The polarization of nuclear spins involves elevated magnetic fields (see below), however, this plot assumes observation in a magnetic field $\approx 10~{\rm G}$ in both cases.  One can also detect the spin state by monitoring transmission of infrared light tuned to the 1042 nm transition between the two singlet levels, resulting in improved light collection \cite{Acosta2010}.
%

\begin{figure}
  \includegraphics[width=3.4 in]{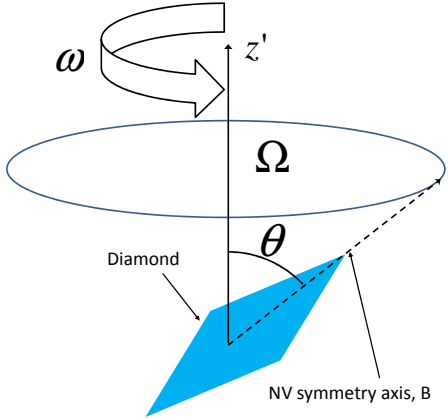}\\
  \caption{Rotation of diamond about an axis $z'$ oriented at an angle $\theta$ with respect to the ${\rm NV^-}$ symmetry axis. }\label{Fig:RotationScheme}
\end{figure}

\textit{NV$^-$ electronic Berry phase based gyroscopes} --
We first neglect ${\rm ^{14}N}$ hyperfine coupling and suppose an ensemble of ${\rm NV^-}$ centers is pumped into the $m_S = 0$ state, with a magnetic field (approx 10 G) applied along the ${\rm NV^{-}}$ axis.  Suppose the diamond is rotated with an angular velocity $\omega$ about an axis $z'$ oriented at an angle $\theta$ relative to the ${\rm NV^-}$ symmetry axis ($\theta \le \pi/2 $) during some interval $t$ (Fig. \ref{Fig:RotationScheme}).  The symmetry axis thus traces a solid angle
\begin{equation}\label{Eq:solidangle}
\Omega = \int_{\phi=0}^{\omega t}\int_{\theta' = 0}^\theta d\phi d\theta' \sin\theta' =  \omega t (1-\cos\theta).
\end{equation}
In the course of this rotation, the Zeeman eigenstates of the ${\rm NV^-}$ center pick up a geometric phase shift $\Delta\phi=m_S\Omega$. From Eq. \eqref{Eq:solidangle}, we see that this geometric phase shift is maximized for $\theta = \pi/2$. If one establishes coherences between $m_S = 0$ and $m_S = \pm 1$ states prior to rotation, the observed transition frequencies will be $\omega_\pm = 2\pi D\pm g_s\mu_B B \pm \omega(1-\cos\theta)$.

The observed frequency shifts associated with rotations involves both the rotation rate $\omega$ and the angle of rotation with respect to the ${\rm NV^-}$ symmetry axis, $\theta$.  If the magnetic field is oriented such that resonances due to all four ${\rm NV^-}$ orientations are resolved, Berry phase shifts in the four different orientations can be used to determine both the direction of rotation and its magnitude.

One can envision several schemes for readout.  Here, we consider a Ramsey scheme, in which ${\rm NV^-}$ centers are initialized in the $m_S = 0$ sublevel by application of 532 nm pump light. A $\pi/2$ microwave pulse is applied to the spins, they are allowed to precess freely for a time $t$, and then a second $\pi/2$ pulse is applied, converting coherences back into populations.  Finally the spin state is read out by monitoring fluorescence while illuminating the sample with 532 nm pump light.  The resulting fluorescence is given by
\begin{equation}
F \approx \eta N(1+R\cos\omega_\pm t).
\end{equation}
Here, $R$ is the detection contrast for states with $m_S = 0$ and $m_S = \pm 1$, $\eta$ is the collection efficiency, and $N$ is the number of participating ${\rm NV^-}$ centers.  This assumes a roughly optimal readout, in which each center emits one photon in the readout stage. 
The slope of the fluorescence with respect to rotation rate $\omega$ (assuming $\omega$ is small) is maximized by tuning $B$ and $t$ such that $(2\pi D + g_s\mu_B B) t = (n\pm 1/4)2\pi$, where $n$ is an integer, thereby optimizing the sensitivity to rotations. Under such conditions $dF/d\omega \approx \pm \eta N Rt$.
The accuracy with which one can determine the rotation frequency $\omega$ is $\delta \omega = (dF/d\omega)^{-1} \delta F$. Fundamentally limiting the noise in the fluorescence is photon shot noise, $\delta F = \sqrt{\eta N}$.  Setting $t = T_2^\star$, and repeating the measurement $\tau/T_2^\star$ times ($\tau$ is the total integration time), we find
\begin{equation}\label{Eq:SQL}
\delta \omega = 1/(R\sqrt{\eta N T_2^\star \tau}).
\end{equation}

%
Realistic parameters may be conservatively estimated based on Ref. \cite{Acosta2009}, where ${\rm NV^-}$ ensembles were realized with density of $n=10^{18}~{\rm cm^{-3}}$, $T_2^\star \approx 300~{\rm ns}$, and a measurement volume of $1~{\rm mm^3}$.  Fluorescence detection suffers from poor contrast ratio and poor detection efficiency (typically $R\approx 0.03$ and $\eta \approx 0.01$).  The detection efficiency can be substantially improved to $\eta\approx 1$ by employing IR absorption \cite{Acosta2010}, or by using the side collection technique demonstrated in Ref. \cite{LeSage2012}, where $\eta \approx 0.5$ was realized.  Assuming $\eta = 0.5$, 1/4 of the ${\rm NV^-}$ centers participating, a measurement volume of $1~{\rm mm^3}$, and a measurement time $\tau = 0.5~{\rm s}$ (corresponding to a bandwidth of 1 Hz), Eq. \eqref{Eq:SQL} yields rotation sensitivity of about $5.4\times 10^{-3}~{\rm rad/s/\sqrt{Hz}}$.  This is below the demonstrated sensitivity of the state-of-the-art sensors mentioned above, however, the active part of the sensor is smaller by several orders of magnitude.

The numbers given above are fairly conservative.  
Reference \cite{Pham2012} reports that CPMG decoupling sequences can be used to extend coherence times by an order of magnitude when ${\rm ^{14}N}$ impurities limit the lifetime.
More recent work has shown coherence lifetimes as long as 300 ms in low ${\rm NV^-}$ density samples in isotopically pure ${\rm ^{12}C}$ diamond  using CPMG decoupling sequence at 77 K \cite{Davidson}.  For the work reported in Ref. \cite{Acosta2009}, dipole-dipole coupling between ${\rm NV^{-}}$ centers appeared to limit the lifetime.  As suggested in Ref. \cite{Taylor2008}, such relaxation may be mitigated by pulse sequences such as MREV \cite{Mansfield1971} or WAHUHA \cite{Wahuha}, which average the dipole-dipole interaction to zero and scale the rank-1 gyroscopic phase shift (for WAHUHA, the scaling factor is $1/\sqrt{3}$). Ultimately, $T_1$ places an upper bound on $T_2$.  In Ref. \cite{Jarmola2012} room-temperature measurements of $T_1 = 3~{\rm ms}$ was measured for the high ${\rm NV^-}$ density samples of present interest.   Assuming dipole-dipole relaxation can be sufficiently suppressed in high-density ensembles, one can expect roughly two orders of magnitude improvement in sensitivity over the estimate given above, placing the sensitivity in the range of $10^{-5}~{\rm rad/s/\sqrt{Hz}}$.
%

Magnetic-field fluctuations present a serious source of noise in this scheme.
The sensitivity to magnetic field fluctuations is given by $\delta\omega = (\partial F/\partial\omega)^{-1}(\partial F/\partial B)\delta B = g_s\mu_B\delta B$.
In order to reach an angular-frequency sensitivity of $5.4\times 10^{-3}~{\rm rad~s^{-1}}$, one requires magnetic field stability $\delta B \approx 3\times 10^{-10}~{\rm G}$.
Assuming $\approx 10~{\rm G}$ is required to sufficiently suppress the stress parameter, this corresponds to fractional field stability at the level of $10^{-10}$, challenging, though not insurmountable.

The temperature dependence of the $D$ coefficient places substantial requirements on thermal stability.  The temperature dependence of $D$ near room temperature is linear \cite{Acosta2010PRL}, with a slope of about -70~kHz/K.  To reach $10^{-3}~{\rm rad/s/\sqrt{Hz}}$ requires thermal stability at the level of $10^{-9}$~K.  Such stability seems unrealistic, however, if one monitors the splitting of the ODMR signal for $m_S = \pm 1$ rather than the position of just a single line, the temperature dependence should be eliminated.  This also improves the sensitivity of the gyroscope by a factor of $\sqrt{2}$.
%


\begin{figure}
  \includegraphics[trim = 0.5in 0.5in 0.5in 0.5in, clip, width=3.4 in]{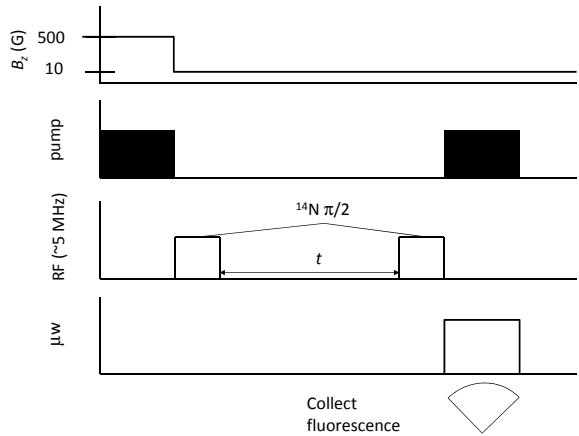}\\
  \caption{Pulse sequence for Berry phase based gyroscope using resolved ${\rm ^{14}N}$ hyperfine structure in diamond.}\label{Fig:14Npulsesequence}
\end{figure}

\textit{Nuclear spin based diamond gyroscopes} --
Nuclear spins are appealing in the context of gyroscopes because their gyromagnetic ratio is about three orders of magnitude smaller than that of the electron, relaxing the requirements on field stability. Recent work has shown that it is possible to achieve very high, $\approx 98\%$, polarization of single ${\rm ^{14}N}$ or ${\rm ^{15}N}$ nuclear spins in the ${\rm NV^-}$ center (as well as nearby ${\rm ^{13}C}$ spins) using an excited-state level anticrossing at about 500 G, induced by the hyperfine coupling of the nucleus with the electrons of the center \cite{Jacques2009,Smeltzer2009}. This technique has been extended to polarization of ${\rm ^{14}N}$ and proximal ${\rm ^{13}C}$ nuclear spins in ensembles of ${\rm NV^-}$ centers \cite{Fisher2012}.  The ${\rm ^{14}N}$ spin state can be detected via fluorescence or IR absorption by the ${\rm NV^-}$ center.

An experimental protocol for a ${\rm ^{14}N}$ diamond based gyroscope is presented in Fig. \ref{Fig:14Npulsesequence}.  ${\rm NV^{-}}$ centers are polarized in the $|m_S = 0,m_I=-1\rangle$ state by turning the magnetic field (oriented parallel to the ${\rm NV^-}$ symmetry axis) up to about 500 G, while illuminating the sample with 532 nm light (see Refs. \cite{Jacques2009,Fisher2012}).  The light is turned off, and the magnetic field reduced to about 10 G (to improve stability), and coherences between the $|m_S = 0,m_I=-1\rangle$ and $|m_S = 0,m_I=0\rangle$ ground state sublevels are established by application of a $\pi/2$ pulse, resonantly tuned to the hyperfine transition frequency $\omega_{hf}\approx 5.1~{\rm MHz}$.  Spins are allowed to evolve freely for some time $t$ in the presence of the magnetic field, hyperfine coupling, and sample rotation, accruing both nuclear Zeeman, hyperfine, and geometric phase shifts.  A second $\pi/2$ pulse converts these coherences back into ${\rm ^{14}N}$ Zeeman-sublevel populations, which are then selectively read out by applying microwaves tuned to the desired ${\rm ^{14}N}$ hyperfine transition, indicated by either of the downwards pointing arrows in Fig. \ref{Fig:14NHyperfine}(b). Population in the selected level would oscillate sinusoidally with a frequency proportional to $\omega_{nuc} = \omega_{hf}+\gamma_N B+\omega$ where $\gamma_N$ is the gyromagnetic ratio of the  nitrogen nucleus (we have set $\theta =\pi/2$ so that the factor $1-\cos\theta$ in Eq. \eqref{Eq:solidangle} is equal to 1).  Since the microwave and optical transitions preserve $m_I$, this results in a sinusoidal modulation of the resulting fluorescence $F$
\begin{equation}
F \propto 1+R \frac{\Delta B^2}{(B-B_0)^2+\Delta B^2} \cos\omega_{nuc}t.
\end{equation}
Here, $B_0$ is the magnetic field corresponding to resonance with the applied microwaves and $\Delta B = 1/(T_2^\star g_s\mu_B)$ is the width of the microwave transition in magnetic field units.  The Lorentzian profile for the microwaves is included to address the possibility of noise due to magnetic field fluctuations. As we see below, it turns out that this factor drops out. The same arguments used to derive the sensitivity, Eq. \eqref{Eq:SQL}, of gyroscopes based on pure electronic transitions applies in this case as well.

The exact value of $T_2^\star$ will depend on conditions and sample preparation.  We point out that in some instances, the nuclear spin coherence times $T_{2,nuc}^\star$ can be significantly longer for nuclear spins than for electron spins. For example, with single ${\rm NV^{-}}$ centers, Ref. \cite{Fuchs2011} reports, $T_{2,e}^\star\approx 1.4~{\rm \mu s}$, while $T_{2,nuc}^\star\approx 58~{\rm \mu s}$ for ${\rm ^{14}N}$ spins. In ${\rm ^{13}C}$ depleted diamonds with low ${\rm NV^-}$ density, ${\rm ^{14}N}$ coherence lifetimes $T^\star_2 = 7~{\rm ms}$ have been realized \cite{Waldherr2012}.  Measurements of these quantities in high-density samples are planned.  Taking $T_{2,nuc}^\star = 1~ {\rm ms}$, equation \eqref{Eq:SQL} yields sensitivity at the level of $9\times 10^{-5}~{\rm rad/s/\sqrt{Hz}}$, with other parameters as used above, and without requiring dynamical decoupling.

Despite the fact that microwave transitions are used to selectively read out the nuclear spin state, the sensitivity to magnetic field fluctuations is dramatically reduced compared to the case of gyroscopes based solely on electronic transitions: $\delta\omega = (\partial F/\partial\omega)^{-1}(\partial F/\partial B)\delta B = \gamma_N\delta B$.  We see that this is suppressed relative to the case of electronic coherences by a factor of $\gamma_N/(g_s\mu_B)\approx 9000$ for ${\rm ^{14}N}$, greatly relaxing the requirements on field stability.

In summary, we have proposed a new type of solid-state gyroscope based on Berry-phase effects in diamond nitrogen-vacancy centers. We estimate that the sensitivity of such a device could surpass that of commercial vibrating MEMS sensors, which find wide application in a variety of technologies.  We also point out that ${\rm NV^-}$ centers have been used to polarize bulk ${\rm ^{13}C}$ in elevated magnetic fields \cite{King2010}, and efforts are under way to produce polarization of bulk ${\rm ^{13}C}$ nuclear spins in low magnetic field.  Bulk ${\rm ^{13}C}$ spins can potentially have very long $T_1$ times \cite{Lefman1994}, and if decoupling schemes can yield correspondingly long $T_2^\star$ times, one may expect dramatic improvements in sensitivity.

This work was supported by AFOSR/DARPA QuASAR, and by the NSF Grant PHY - 0855552. K.J. acknowledges support from the Danish Council for Independent Research. The authors thank V. M. Acosta and A.O. Sushkov for invaluable discussions.


\end{document}